\documentclass[manuscript]{acmart}

\usepackage{color, soul}
\usepackage{algorithm, algorithmic}
\usepackage{hyperref}
\hypersetup{colorlinks = true,
            linkcolor = blue,
            urlcolor  = blue,
            citecolor = blue,
            anchorcolor = blue}
\AtBeginDocument{%
  \providecommand\BibTeX{{%
    \normalfont B\kern-0.5em{\scshape i\kern-0.25em b}\kern-0.8em\TeX}}}

\renewcommand\footnotetextcopyrightpermission[1]{}
\titlenote{Copyright 2021 for this paper by its authors. Use permitted under Creative Commons License Attribution 4.0 International (CC BY 4.0).\\Presented at the MORS workshop held in conjunction with the 15th ACM Conference on Recommender Systems (RecSys), 2021, in Amsterdam, Netherlands.}

\begin{document}

\title{
Evaluating Music Recommendations with Binary Feedback for Multiple Stakeholders
}


\author{Sasha Stoikov}
\affiliation{%
\institution{Cornell Financial Engineering, Cornell Tech}
\city{New York}
\country{USA}}
\email{sfs33@cornell.edu}

\author{Hongyi Wen}
\affiliation{%
\institution{Cornell Information Science, Cornell Tech}
\city{New York}
\country{USA}}
\email{hw557@cornell.edu}

\thanks{Both authors contributed equally to this research.}


\begin{abstract}
High quality user feedback data is essential to training and evaluating a successful music recommendation system, particularly one that has to balance the needs of multiple stakeholders. Most existing music datasets suffer from noisy feedback and self-selection biases inherent in the data collected by music platforms. 
Using the Piki Music dataset of 500k ratings collected over a two-year time period, we evaluate the performance of classic recommendation algorithms on three important stakeholders: \textit{consumers, well-known artists and lesser-known artists}. 
We show that a matrix factorization algorithm trained on both likes and dislikes performs significantly better compared to one trained only on likes for all three stakeholders. 
\end{abstract}

\keywords{Datasets, Music recommendations, Multi-stakeholders}

\maketitle

\section{Introduction}


Music recommendation algorithms play a major role in what music gets listened to on streaming platforms~\cite{mehrotra2020bandit, millecamp2018controlling}. This in turn influences which artists make a living from streaming and which ones do not. Understanding the mechanisms that cause an algorithm to push certain artists ahead of others is increasingly urgent. In this paper, we focus on three stakeholders of a music recommendation system: \textit{music consumers, well-known artists and lesser-known artists}.

Much has been written about competing classes of algorithms and metrics. Unfortunately, the growing number of proposed metrics makes it difficult to answer straightforward questions related to the satisfaction of the stakeholders of a music streaming platform. For example, what proportion of recommendations is actually liked by users? Do the algorithms serve well-known artists better than lesser-known artists? Are music consumers more or less satisfied with well-known or lesser-known recommendations? There is often a tension between metrics that aim to measure familiarity, relevance and predictability, and metrics that aim to measure fairness, diversity and serendipity. Combining these metrics into objective functions that represent the interests of the aforementioned stakeholders is challenging. 

At the heart of this problem are three common assumptions in the data used for training and testing recommendation algorithms: (i) unheard songs are assumed to be disliked, though in reality unheard songs are often excellent and (ii) played songs are assumed to be liked, though they may have been passively listened to on a playlist (iii) rated songs are randomly presented to users, while they are often self-selected by the user. All of these assumptions tend to favor the well-known artists, who are more often heard, more often recommended on playlists and more often remembered at the search bar. To mitigate these concerns we built Piki~\footnote{\href{www.piki.nyc}{www.piki.nyc}}, a music discovery tool that collects ratings through a binary choice (like/dislike) on song clips that cannot be searched or skipped. We present the  \textit{Piki Music} dataset with the goal of enabling researchers and practitioners from the RecSys community to explore the above challenges. We highlight the following contributions:
\begin{itemize}
    \item We collect the Piki Music dataset in a way that incentivizes users to give truthful binary ratings to songs, thus mitigating the self-selection and noisy feedback biases inherent in most recommendation datasets.
    (Section 3).
    \item We quantify the value of a dislike by training a matrix factorization algorithm on this binary dataset. We find that it performs significantly better than when trained on positive-only feedback for all three stakeholders (Section 4).
\end{itemize}

\section{Related work}
\label{sec:related-work}
\paragraph{\textbf{Datasets with Explicit and Implicit User feedback}}

Datasets consisting of user-item interactions are typically elicited in two ways: explicitly or implicitly. Explicit feedback refers to an action that a user performs with the intention of giving their opinion on the quality of an item, for example, giving a rating to a watched movie. Two well-known datasets for recommendations, the Netflix Prize dataset~\cite{bennett2007netflix} and the Movielens dataset~\cite{movielens}, are examples of explicit feedback datasets. Both consist of 1-5 star ratings for millions of user-item pairs. Explicit feedback is voluntary, so this kind of dataset is often subject to self-selection bias: if there is no obligation to rate, users tend to rate only when they feel very strongly about an item, and rate more items they like than items they dislike. Unbalanced explicit training datasets have been documented in the literature and some flattening techniques have been shown to improve recommendation quality~\cite{mansoury2021flatter}. 

Implicit feedback refers to an action that a user performs with an intention other than giving their opinion on the quality of the item, such as clicks, purchases, etc. Implicit feedback collected by streaming apps is commonly used to train modern recommendation systems~\cite{10.1109/ICDM.2008.22}. However, researchers have expressed concerns about training and evaluating algorithms using this type of feedback data~\cite{10.1145/3298689.3347037,10.1145/3209978.3210007}. A major problem with implicit feedback datasets for music recommendations, such as those obtained on Lastfm~\cite{bertin2011million} or Spotify~\cite{brost2019music}, is that they may measure a noisy signal of the user's true preferences. If a user listens to a song, that action may be interpreted as positive feedback, though the song may have been passively played as background music~\cite{10.1145/3298689.3347037}. 
Moreover, implicit feedback typically consists of ``positive-only'' data. To train and test an algorithm, missing user-item interactions are typically labeled as negative feedback, which only adds more noise to the training data. One approach to deal with implicit data is to label songs that are less often played as negative feedback ~\cite{Sanchez}. However, few datasets with explicit positive and negative user preferences are publicly available, making it hard to investigate the limitations of implicit feedback.

\paragraph{\textbf{Recommendations for multi-stakeholders}}

A recent paper calls for a paradigm shift from user-centric metrics towards modeling the value that recommendation systems bring to their multiple stakeholders~\cite{jannach2020escaping}. They identify potential stakeholders as consumers, producers, platforms and society at large and call for evaluation designs aligned with the goals of each stakeholder. The way forward, they suggest, is to design better evaluation methods that combine multiple goals and generalize beyond domain specific applications. Researchers have investigated recommendations from the multi-stakeholder perspective by optimizing over multiple objectives such as diversity, relevance, fairness, satisfaction~\cite{mehrotra2020advances, mehrotra2020bandit}. They highlight an inherent tension between relevant recommendations that are similar to past consumption and diverse recommendations that are outside of a user's echo chamber. Fairness in compensating for the popularity bias has gathered particular attention~\cite{abdollahpouri2020unfair,kowald2019unfairness, Celma}. The authors suggest addressing multiple stakeholders by modeling profit-aware recommendations ~\cite{abdollahpouri2019beyond, abdollahpouri2017multiple}, e.g., recommendations that are linked to sales. However, a challenge is the lack of publicly-available data with multi-stakeholder characteristics, which tends to be sensitive data.

In the music domain, lesser-known artist have expressed many concerns, which include reaching an audience, transparency in recommendations, localizing discovery, gender balance and popularity bias, according to a qualitative study ~\cite{ferraro2021fair}. 
In the sequel, we take a quantitative approach to some of these concerns with the Piki Music Dataset.

\section{Piki Music Dataset}

Through the Piki interface (Fig. ~\ref{fig:interface}), we collect binary data while incentivizing users to provide feedback in a way that is aligned with their individual tastes. The Piki Music dataset currently consists of 2723 anonymized users, 66,532 anonymized songs and 500K binary ratings and the data collection is on-going. Figure~\ref{fig:data-distribution} illustrates the distribution of like rates  across users and songs. The columns of the dataset are as following: 

\begin{figure}[t]
        \includegraphics[width=90pt]{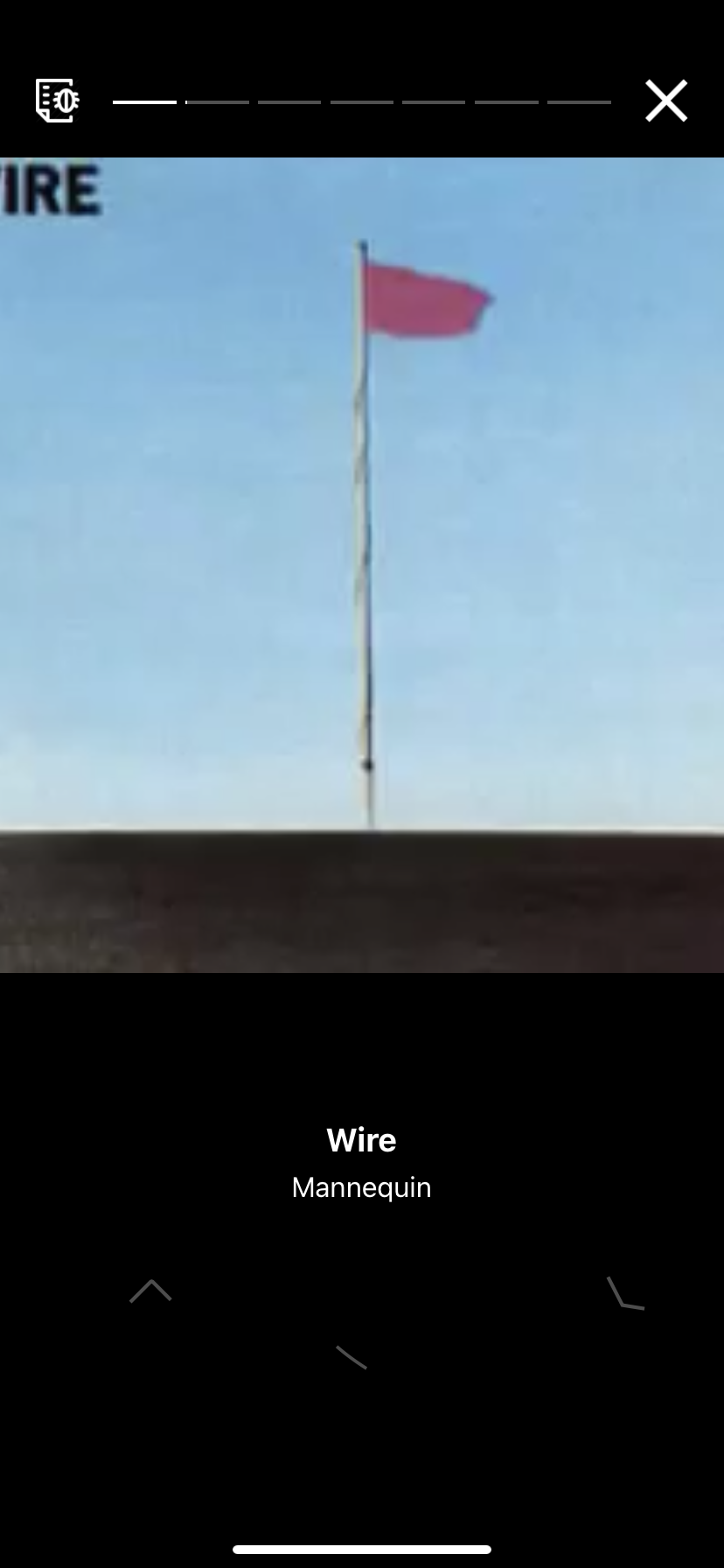}
        \includegraphics[width=90pt]{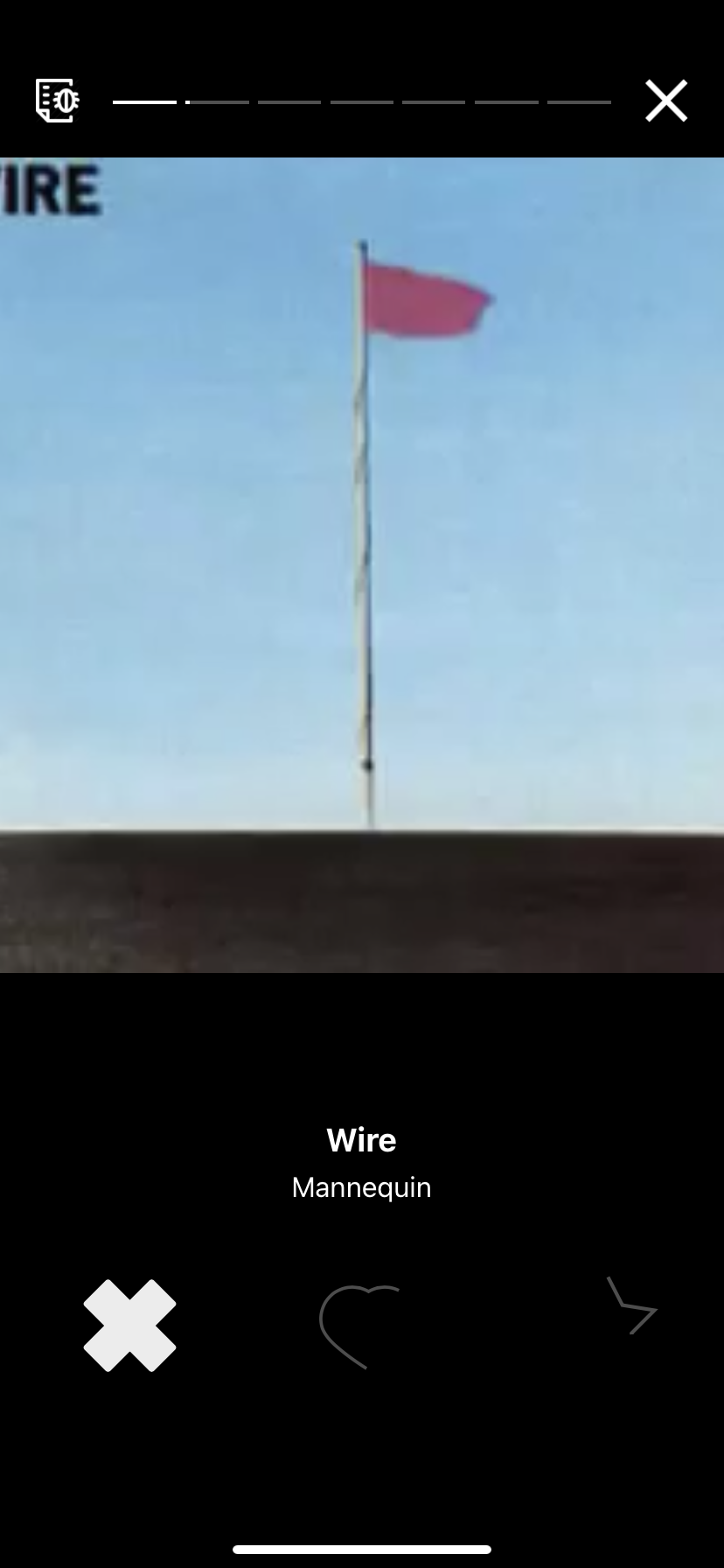}
        \includegraphics[width=90pt]{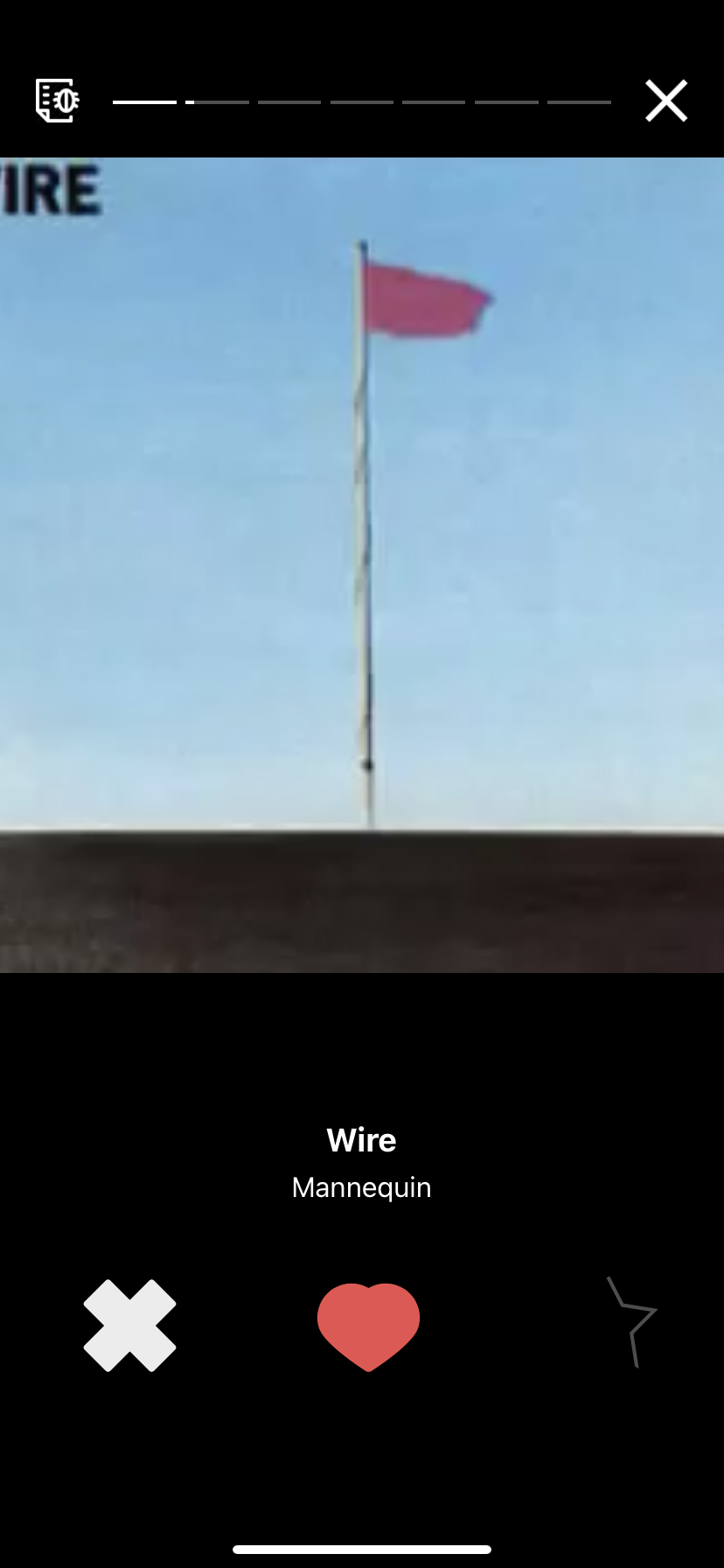}
        \includegraphics[width=90pt]{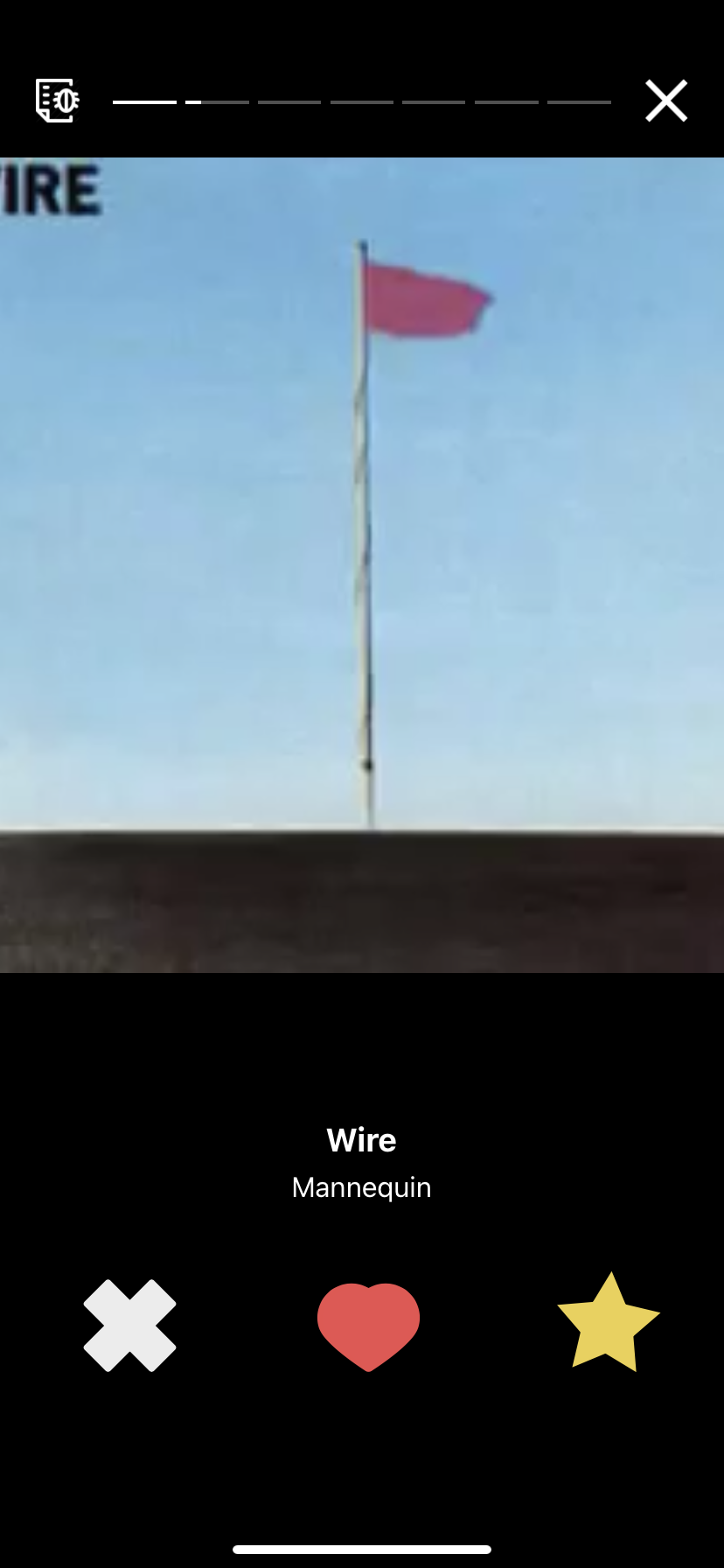}
    \caption{The Piki Music App interface. The dislike button is unlocked after 3 seconds, the like button after 6 seconds and the superlike button after 12 seconds. This incentives the users to vote truthfully: to dislike is easy but in order to like, a user must invest time in the song.}
    \label{fig:interface}
\end{figure}
 
 \begin{figure}[!t]
        \includegraphics[width=180pt]{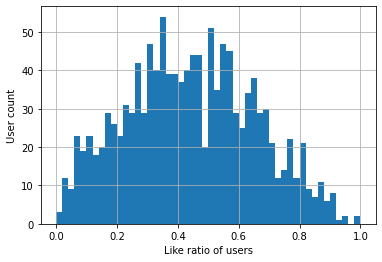}
        \includegraphics[width=180pt]{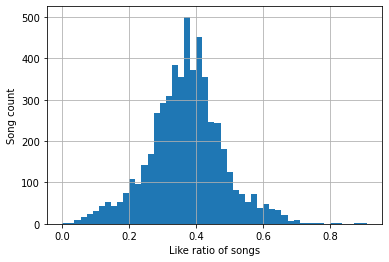}
    \caption{The distribution of like ratios across users and songs.}
    \label{fig:data-distribution}
\end{figure}

\begin{itemize}
    \item timestamp: a datetime variable
    \item user\_id: an anonymized user id
    \item song\_id: an anonymized song id
    \item liked: this is the binary indicator, 1 if the song is liked, or 0 if the song is disliked. Note that the feedback consists of 39\% likes and 61\% dislikes. The superlike indicator, labeled 2, is included in the data, though we treat it as a like in our experiments.
    \item personalized: this is 1 if the song was recommended based on their previous choices or 0 if the song was selected randomly. Note that the songs recommended are 66\% personalized and 34\% random songs. We have included this flag in the dataset, to allow mitigation of the recommendation bias of the data, though this question is not within the scope of our study.
    \item spotify\_popularity: this is the song's artist's popularity, a value between 0 and 100, with 100 being the most popular. It is published by Spotify for each artist, through their publicly-available API~\footnote{\href{https://developer.spotify.com/documentation/web-api/reference/\#category-artists}{https://developer.spotify.com/documentation/web-api/reference/category-artists}}. The average value of the Spotify popularity in our data set is 52, so we classify songs as coming from well-known artists if the value is above this mean and as a lesser-known artist if it is below the mean. Note that this threshold corresponds to artists that have approximately 350,000 monthly listeners, which on average generates around \$2000 per month, assuming this is a solo artist without a label.
\end{itemize}

\subsection{Binary data collection}

Users on Piki provide explicit binary feedback, by liking or disliking music clips. Users do not have access to a search bar and thus cannot control what songs they will hear. They are asked to like or dislike 30 second music video clips. Figure 1 shows how the interface presents the songs in batches, much like a social media story.  The binary nature of the Piki music data set addresses our first concern with the training data, namely, that we won't need to treat unheard songs as disliked, since we have a set of disliked songs to train and test the algorithm.
 
 \subsection{Incentives to vote truthfully}
 
 The dislike button is unlocked after 3 seconds (see the first and second images in Fig.~\ref{fig:interface}), the like button is unlocked after 6 seconds (see the 3rd image in Fig.~\ref{fig:interface}) and the superlike button is unlocked after 12 seconds (see the 4th image in Fig.~\ref{fig:interface}). The clip starts 40 seconds into the song and when the clip is over, 30 seconds later, the user may replay the clip before rating it. This mechanism aligns the users' ratings with their preferences. The 3 second lock period for the dislike button ensures that all songs are given a fair chance. The 6 and 12 second lock periods guarantee that positive ratings are backed up by a meaningful time investment in the song. Only songs that truly capture a user's attention get liked or superliked. Piki users are rewarded with micropayments each time they complete a set of ratings. Both the timing and the financial rewards help mitigate our second data concern, namely that liked songs are actively liked, not just played passively on a playlist.

\section{Experiments}
We split the dataset into a training set $\mathcal{T}$ and an evaluation set $\mathcal{E}$ using random sampling according to 80\%/20\% splits stratified by user and average the results across 5 runs. Algorithms are trained on $\mathcal{T}$ and scores are computed on interactions in $\mathcal{E}$. 

\subsection{Training matrix factorization algorithms}

Given a set of users $\mathcal{U}$, songs $\mathcal{I}$ and ratings $\mathcal{R}$, collaborative filtering aims to learn $d$-dimension latent user vectors $x \in \mathbb{R}^{|\mathcal{U}| \times d}$ and latent item vectors $y \in \mathbb{R}^{|\mathcal{I}|\times d}$ from the sparse user-item rating matrix through singular value decomposition (SVD)~\cite{paterek2007improving}. Predicted user preference scores are given by the dot products between user and item vectors:
\begin{equation}
\label{eq:mf-pred}
\hat{p}_{ui} = x_{u}^Ty_{i} .
\end{equation}

A classic algorithm from this framework is the Weighted Regularized Matrix Factorization (WRMF)~\cite{10.1109/ICDM.2008.22}. Specifically, WRMF optimizes for the following objective:
\begin{equation}
\label{eq:wrmf}
\min_{x^*y^*} \sum_{u,i}c_{ui}(r_{ui} - \hat{p}_{ui})^2 + \lambda R(x,y) , 
\end{equation}
where $r_{ui}$ is the ground true preference score for user $u$ to item $i$, $c_{ui}$ is the weight put on each observation, $R(x,y) = ||x||_f
^2 + ||y||_f^2$ represents the Frobenius norms for regularizing user and items matrices, and $\lambda$ is regularization parameter. Note that with implicit feedback datasets, $r_{ui}$ is assumed to be $\{0,1\}$, where a click is treated as 1 and missing data from the rating matrix is treated as 0. 

In a setting where negative feedback is available, a generalized framework is proposed to incorporate negative feedback during training~\cite{10.1145/3298689.3347037}. The objective function can be written as:

\begin{equation}
\label{eq:wrmf-ext}
\min_{x^*y^*} \alpha\sum_{(u,i)\in \mathcal{T}_{P}}(1-\hat{p}_{ui})^2  
+ \beta \sum_{(u,i)\in \mathcal{T}_{N}}\hat{p}_{ui}^2  
+ \gamma \sum_{(u,i)\in \mathcal{T}_{M}}\hat{p}_{ui}^2
+ \lambda R(x,y),
\end{equation}
where $\mathcal{T}_P, \mathcal{T}_{N}, \mathcal{T}_{M}$ refer to positive, negative and missing user feedback, $\alpha,\beta,\gamma$ are weights assigned to the corresponding sets of user feedback. We highlight two types of weight schemas:
\begin{itemize}
    \item \textbf{WRMF with Likes}: $\alpha=\gamma=0.5$, $\beta=0$. This means that we only sample from the positive and missing feedback, while ignoring the negative feedback.
    \item \textbf{WRMF with Likes and Dislikes}: $\alpha=\beta=0.5$, $\gamma=0$. This means that we only sample from positive and negative feedback, without the need to sample from missing data.
\end{itemize}

\subsection{Implementations}
We implemented the WRMF algorithm based on the OpenRec library~\cite{yang2018openrec}. We used the Adagrad optimizer with a learning rate of 0.01 and a batch size of 512 to train the model. The regularization parameter $\lambda$ is tuned on a validation set from $\{0.1, 0.01, 0.001, 0.0001\}$ and early stopping is performed. We used a dimension of $d=20$ for both user and item latent factors. The code for experiments and the Piki Music Dataset is public~\footnote{\href{https://github.com/sstoikov/piki-music-dataset}{https://github.com/sstoikov/piki-music-dataset}}. 

\subsection{Performance metrics}
\label{sec:define-metric}

The evaluation dataset $\mathcal{E}$ can be segmented into
 $\mathcal{E}_P, \mathcal{E}_{N}, \mathcal{E}_{M}$ referring to positive, negative and missing user feedback. Popular metrics like Recall@$k$ and Precision@$k$ aim to quantify the relevance of playlists of length $k$. However, they implicitly assume that $\mathcal{E}_{N}$ and  $\mathcal{E}_{M}$ are indistinguishable from each other. 
 
 For a trained algorithm, all scores above a given threshold are classified as recommendations.  Without loss of generality, we use the median of scores from model outputs as the threshold in our experiments. We use $\mathcal{R}$ to denote the songs recommended by an algorithm from $\mathcal{E}$. The set of recommended songs from the well-known artists is $\mathcal{R}_w$ and the set of recommended songs from the lesser-known artists is $\mathcal{R}_l$. It is obvious that $\mathcal{R}= \mathcal{R}_w \cup \mathcal{R}_l$. For each song in $\mathcal{R}$, we have a corresponding binary rating of $\{0,1\}$ from the consumers. For simplicity, we use $S_\mathcal{R}$ to represent the vector for binary ratings on songs in $\mathcal{R}$. Similarly, we have $S_{\mathcal{R}_w}$ and $S_{\mathcal{R}_l}$ .

\begin{itemize}
    \item \textbf{Consumers}: the proportion of the recommended songs from the evaluation set that are actually liked by users: $\frac{\sum S_\mathcal{R}}{|S_\mathcal{R}|}$. The intuition is that a higher precision is aligned with better user experience for the consumers.
    \item \textbf{Well-known artists}: the proportion of recommendations coming from the well-known artists that are actually liked by users:$\frac{\sum S_{\mathcal{R}_w}}{|S_{\mathcal{R}_w}|}$. A higher precision leads to more effective song exposure for well-known artists.
    \item \textbf{Lesser-known artists}: the proportion of recommendations coming from the lesser-known artists that are actually liked by users:$\frac{\sum S_{\mathcal{R}_l}}{|S_{\mathcal{R}_l}|}$. A higher precision leads to more effective song exposure for lesser-known artists.
\end{itemize}

\begin{table}[!t]
    \centering
    \begin{tabular}{l|c|c|c}
        \toprule
         Model & Well-known artists(\%) & Lesser known artists (\%) & Consumers (\%)\\ \hline
         \textbf{Popularity}(recommend well-known)  & $41.1\pm0.04$ & - &  $41.1\pm0.04$ \\ 
         \textbf{Anti-popularity}(recommend lesser-known) & - & $36.3\pm0.06$ & $36.3\pm0.06$ \\ \hline
         \textbf{WRMF with Likes}  & $51.9\pm0.29$ & $48.2\pm0.35$ & $50.1\pm0.28$ \\ 
         \textbf{WRMF with Likes and Dislikes} & $61.3\pm0.18$ & $57.6\pm0.46$ & $59.6\pm0.30$ \\ 
         \bottomrule
    \end{tabular}
    \caption{Performance measured by precision for different stakeholders under WRMF models. }
    \label{tab:result-stakeholders}
\end{table}

\subsection{Results}

We evaluate the performance of the three stakeholders on the following baselines:
\begin{itemize}
    \item \textbf{Popularity}: A naive baseline that always recommend more popular songs from well-known artists, i.e., $\mathcal{R}_l = \emptyset$. 
    \item \textbf{Anti-popularity}: A naive baseline that always recommend less popular songs from lesser-known artists, i.e., $\mathcal{R}_w = \emptyset$.
    \item \textbf{WRMF with Likes}: A matrix factorization algorithm trained on likes~\cite{10.1109/ICDM.2008.22}. It makes the assumption that missing data are negatives. 
    \item \textbf{WRMF with Likes and Dislikes}: A matrix factorization algorithms trained on likes and dislikes using the framework proposed in~\cite{10.1145/3298689.3347037}. 
\end{itemize} 

The Popularity recommender achieves a precision of 41.1\% for consumers, while the Anti-popularity recommender results in a lower precision of 36.3\%. As a comparison, \textit{WRMF with likes} improve the consumer metric by 21.9\% (table~\ref{tab:result-stakeholders}). Moreover, we find that \textit{WRMF with Likes and Dislikes} outperforms the \textit{WRMF with Likes} for all three stakeholders. The consumer metric was lifted by 18.9\%, with an increase of 18.1\% for well-known artists and 19.5\% for lesser-known artists. This highlights the importance of binary feedback in improving the training and evaluation of recommenders.

\section{Conclusion}
\label{sec:future-work}

We present the Piki Music Dataset and argue that it was collected in a way that addresses many of the biases of other publicly available datasets. More importantly, since the ratings are binary (in the form of likes and dislikes), we can define performance metrics for recommendation algorithms from the perspective of various stakeholders. 

There are a few directions that we think future researchers using this dataset may want to explore. In the spirit of the Netflix challenge, we encourage researchers to test the accuracy of more advanced RecSys algorithms on the dataset. For example, it would be interesting to determine if a neural recommender performs better than matrix factorization for consumers, well-known artists or lesser-known artists. 
It may also be valuable to explore other metrics to measure the interests of the stakeholders in this study. Researchers may be also be interested in other ways to segment the artist stakeholders, across genres or other metadata associated with the songs. We are particularly interested in modeling how the algorithm's objectives are tied to the business objectives of other important stakeholders such as streaming platforms and record labels.

\bibliographystyle{ACM-Reference-Format}
\bibliography{main_ref}


\end{document}